\documentstyle[aps,prl,floats,epsf,psfig]{revtex}

\newcommand {\be} {\begin{equation}} 
\newcommand {\ee} {\end{equation}} 
\newcommand {\Be}{\begin{eqnarray*}}
\newcommand {\Ee} {\end{eqnarray*}}
\newcommand {\bey} {\begin{eqnarray}} 
\newcommand {\eey} {\end{eqnarray}} 

\begin{document}
\renewcommand{\theparagraph}{\Alph{paragraph}}
\draft
\twocolumn[\hsize\textwidth\columnwidth\hsize\csname@twocolumnfalse%
\endcsname

\title{
Finite thermal conductivity in 1d lattices
\\}

\author{C. Giardin\'a$^{1,2}$, R. Livi$^{3,4}$, 
        A. Politi$^{5,4}$, and M. Vassalli$^{3}$\\[-2mm]$ $}

\address{
$^{1}$
Dipartimento di Fisica, via Irnerio 46, 40126 Bologna Italy \\
$^{2}$
Istituto Nazionale di Fisica Nucleare, Sez. di Bologna,\\
$^{3}$
Dipartimento di Fisica, L.go E. Fermi 2, 50125 Firenze, Italy\\
$^{4}$
Istituto Nazionale di Fisica della Materia, Unit\`a di Firenze,\\
$^{5}$
Istituto Nazionale di Ottica, L.go E. Fermi 6, 50125 Firenze, Italy\\
}

\date{\today}

\maketitle
\begin{abstract}
We discuss the thermal conductivity of a chain of coupled rotators,
showing that it is the first example of a 1d nonlinear lattice exhibiting 
normal transport properties in the absence of an on-site potential. 
Numerical estimates obtained by simulating a chain in contact with two 
thermal baths at different temperatures are found to be consistent with 
those ones based on linear response theory. The dynamics of the Fourier 
modes provides direct evidence of energy diffusion. The finiteness of the 
conductivity is traced back to the occurrence of phase-jumps. Our 
conclusions are confirmed by the analysis of two variants of this model.
\end{abstract} \vspace{2mm}
\pacs{PACS numbers: }]
%
%
%
%
The understanding of heat conduction in insulating solids is a long-standing 
problem even in the simple context of 1d systems. Rigorous studies have 
shown that the thermal conductivity $\kappa$ diverges (in the thermodynamic 
limit) in homogeneous \cite{Lebowitz} and disordered \cite{Keller} harmonic 
chains. Accordingly, it has become transparent that the nonlinear character 
of the microscopic dynamics is a necessary ingredient for the emergence of 
the Fourier heat conduction law at the macroscopic level.

The typical Hamiltonians chosen to study this problem belong to the general
class
\begin{equation}
\label{model}
H = \sum_{i=1}^{N}\left [ {p_i^2\over 2m_i} + V( q_{i+1} - q_{i}) + U(q_i) 
\right ] ,
\end{equation}
where $m_i$ represents the mass of the $i$-th particle \cite{note1}, $V$ 
is the potential energy of internal forces and $U$ is an on-site potential 
representing interactions with a possible external substrate.

The first convincing numerical evidence of a finite conductivity has been 
provided by the preliminary study of the ding-a-ling model \cite{Casati} 
and the successive detailed analysis of the modified ding-dong model
\cite{Prosen}. Such results have supported the conjecture that a strong
chaotic dynamics is the crucial requisite for the observation of normal
transport properties in 1d chains.  {\sl A fortiori}, this has been assumed 
to hold in higher-dimensions, where scattering effects, due to nonlinear 
interactions among normal modes, should make the diffusion process, at the 
basis of standard transport phenomena, even more efficient than in 1d systems.

However, the recent study of the more realistic FPU $\beta$-model 
($V(x) = x^2/2 + x^4/4$, $U=0$) has revealed a power-law 
divergence of the thermal conductivity as $\kappa \sim N^{2/5}$ \cite{Lepri1}
(where $N$ is the chain length \cite{note2}).
Such an anamalous behaviour has been explained by invoking the 
self-consistent mode coupling theory \cite{Resis} in the description of 
the effective evolution of long-wavelength modes \cite{Lepri2}.
The generality of this result has been confirmed by the study of several 
models with no on-site potential, including the diatomic Toda
model \cite{Hatano} that was previously believed to yield a finite
conductivity. On the other hand, numerical results have accumulated 
indicating that whenever an external force is added (as in the above
mentioned ding-a-ling and ding-dong models), the conductivity is 
finite \cite{Hu}. As a consequence, it has been suggested that momentum 
conservation is the key ingredient responsible for an
anomalous transport even in the presence of a chaotic evolution. 

Such a conjecture has been very recently put on more firm grounds
by Prosen and Campbell \cite{PC99}, who have shown that
anomalous transport necessarily arises whenever two conditions are fulfilled:
({\it i}) the momentum is conserved; ({\it ii)} the pressure does not vanish 
in the thermodynamic limit. On the other hand, systems with symmetric 
potentials, such as  the FPU-$\beta$ model, do not belong to this class, since
the pressure vanishes. In practice, as all momentum-conserving 
models studied in the literature have revealed anomalous transport properties, 
the possibility that the condition on the pressure is redundant remains open.

In this Letter, we show that this is not the case by analysing some 1d 
models where momentum is conserved and yet the conductivity is finite. 

\paragraph{The rotator model}
The simplest example of a classical-spin 1d model with nearest neighbour 
interactions lies in the class (\ref{model}) with $V(x) = 1 - \cos(x)$ 
and $U = 0$.  This model can be read also as a chain of $N$ coupled pendula, 
where the $p_i$'s and the $q_i$'s represent 
action-angle variables, respectively. It has been extensively studied 
\cite{Benettin,Livi} as an example of chaotic dynamical system \cite{Pettini}
that becomes integrable both in the small and high energy limits, when it
reduces to a harmonic chain and free-rotators, respectively. In the two 
integrable limits, the relaxation to equilibrium slows down very rapidly for 
most of the observables of thermodynamic interest (e.g., the specific heat) 
\cite{Livi}. As as consequence, the equivalence between ensemble and time 
averages is established over accessible time scales only inside a limited 
interval of the energy density $\varepsilon$. Here we shall discuss heat 
conduction for values of the energy density corresponding to strongly 
chaotic behaviour. 
\paragraph
{Numerical analysis of the thermal conductivity}
The most natural and direct way to determine $\kappa$ consists 
in simulating a real experiment, by coupling the left and right extrema of
the chain with two thermal baths at temperatures $T_{L} > T_{R}$, 
respectively. In our simulations we have used Nos\'e-Hoover models of 
thermostats \cite{Nose}, both because they can be easily
implemented (integrating the resulting equations with a standard algorithm)
and because of the smaller finite-size effects (due to the unavoidable 
contact resistance). 

With this setting, a non-equilibrium stationary state sets in characterized
by a non-vanishing heat flux $J$ \cite{Choquard,Lepri1}:
\begin{equation}
J \;=\; {1\over N} \sum_{i} j_{i}
  \;=\; {1\over N} \sum_{i}  {p_{i}\over 2}\, \left(f_{i+1}+f_{i}\right)
\end{equation}
\noindent
where $f_{i}= -{\partial V(q_{i+1}-q_i)}/{\partial q_i}
= \sin (q_{i+1} - q_{i})$ is the interaction force and $j_i$ is the 
local flux at site $i$. The total heat flux $J$ has to be averaged over a 
sufficiently long time span to get rid of fluctuations and to ensure the 
convergence to the stationary regime. This can be tested by monitoring the 
average heat flux and looking at the scale of its fluctuations. 
As a result, we have verified that $2\cdot 10^6$ time units are sufficient 
to guarantee a few percents of fluctuations in the worst cases.

\begin{figure}[h]
\psfig{file=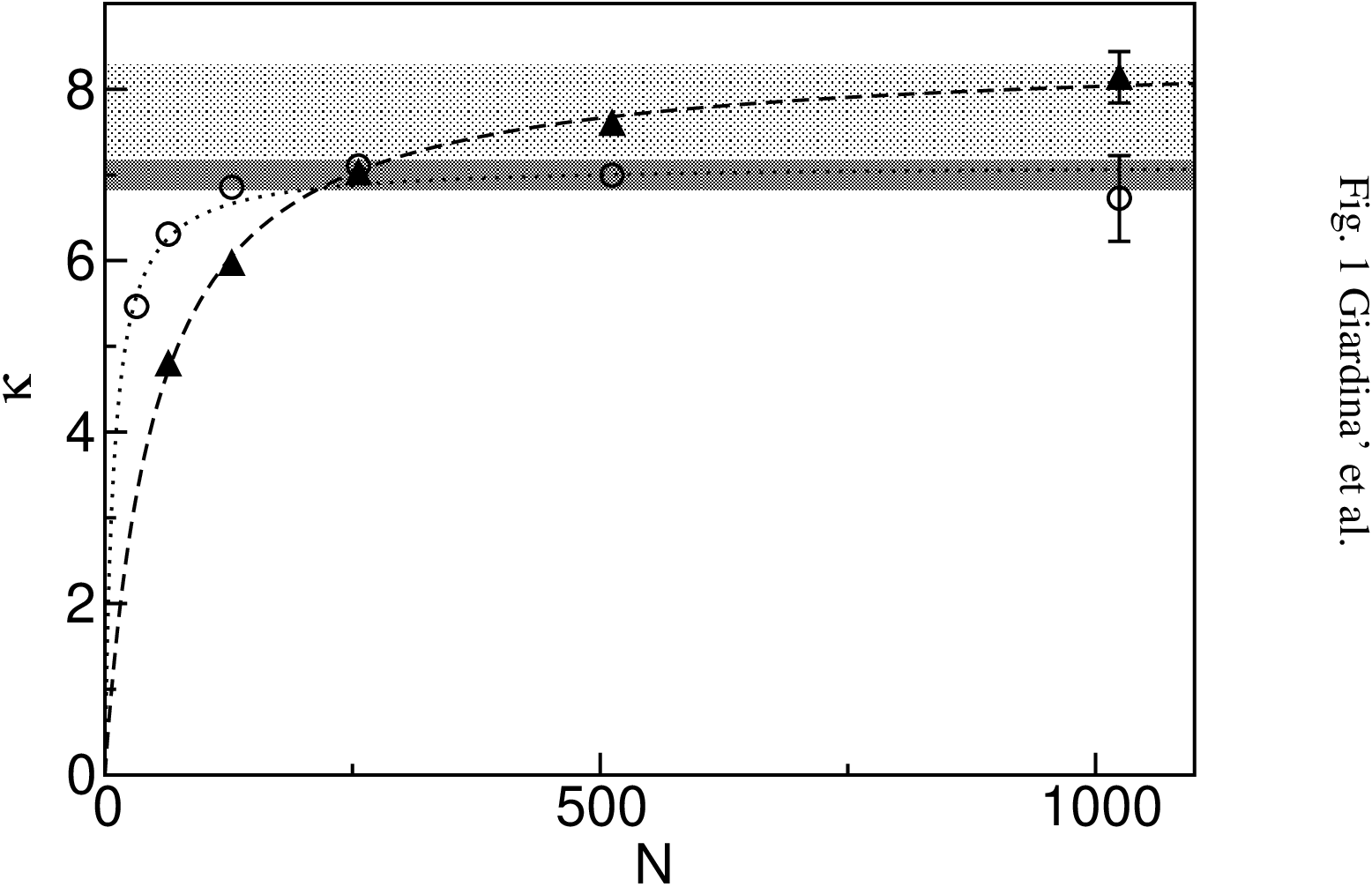,width=8.5 truecm,angle=0}
\caption{Conductivity $\kappa$ versus chain length $N$ as
obtained from non-equilibrium molecular dynamics. Circles correspond
to the rotator model with temperatures $T_L = 0.55$ and $T_R = 0.35$;
triangles correspond to the double well potential with $T_L = 0.04$
and $T_R = 0.06$. The two lines represent the best fit with 
the function $a + b/N$. The two shaded regions represent the uncertainty
about the conductivity on the basis of the Green-Kubo formula.}
\label{av_flux}
\end{figure}

The thermal conductivity is determined by assuming the Fourier law, 
i.e. from the relation  $J = \kappa \nabla T$,
where $\nabla T$ denotes the imposed thermal gradient. The simulations have
been performed for $T_L = 0.55$, $T_R= 0.35$, and chain lengths ranging from
$N=32$ to 1024 with fixed boundary conditions. The equations of 
motion have been integrated with a 4th-order Runge-Kutta algorithm and a
time step $ \Delta t = 0.01$. The results, reported in Fig.~\ref{av_flux} 
clearly reveal a convergence to a $\kappa$-value approximately equal to 7
(see the circles). The dotted line represents the best data fit with the   
function $a + b/N$: the agreement is very good, showing that finite-size
corrections to $\kappa$ are of the order $O(1/N)$, as it should 
be expected because of the thermal contacts. However, more important 
than the numerical value of the conductivity is its finiteness in spite 
of the momentum conservation.

In order to test independently the correctness of our results, we have 
performed direct microcanonical simulations, which allow determining the 
thermal conductivity through the Green-Kubo formula \cite{Green}:
\be
\kappa = {1\over T^2} \int_0^{\infty} C_J (t)dt
\label{GK}
\ee
where $C_J (t)=N\langle J(t)J(0)\rangle$ is the flux autocorrelation
function at equilibrium and $T$ is the temperature. A correct
application of the above formula requires fixing the energy density 
$\varepsilon$ in such a way that the kinetic temperature (defined as 
$T = \langle p^2 \rangle$, in agreement with the virial theorem) is close
to the average value of the temperature in the previous simulations. 
The choice $\varepsilon = 0.5$ turns out to be reasonable, as it corresponds 
to $T \approx 0.46$. In the absence of thermal baths, the equations of 
motion are symplectic, so that we have now preferred to use a 6th order 
McLachlan-Atela integration scheme \cite{Mclachlan} (with periodic boundary
conditions).

\begin{figure}[h]
\psfig{file=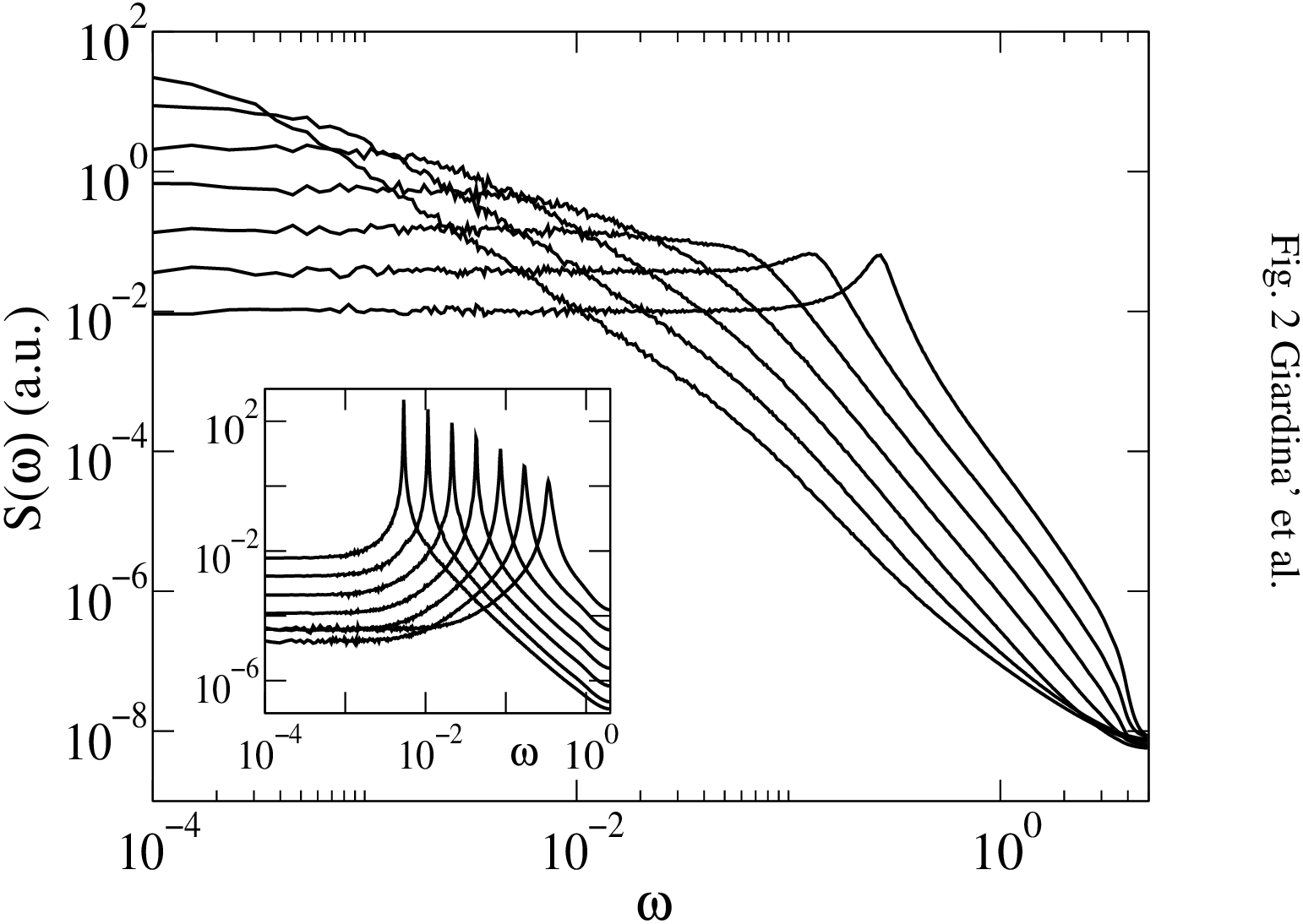,width=8.5 truecm,angle=0}
\caption{Power spectra $S(w)$ in arbitrary units of the Fourier modes
1,2,4,8,16,32 and 64 (curves from left to right in the high frequency 
`region) in a chain of length $N=1024$ of rotators with energy density
with energy density 0.5. The curves result from an average over 1000 
independent initial conditions. In the inset, the same modes are 
reported for a diatomic FPU chain with masses 1, 2 and energy density 8.8.}
\label{modes}
\end{figure}

The correlation function has been computed by exploiting the Wiener-Khinchin 
theorem, i.e. by anti-transforming the Fourier power spectrum. The result
of the time integration is almost independent of $N$ for $N>128$. The
gray region in Fig.~\ref{av_flux} corresponds to the expected value of $\kappa$
taking into account statistical fluctuations. There is not only a clear
confirmation of a finite conductivity, but the numerical value obtained 
with this technique is in close agreement with the direct estimates.
\paragraph
{Dynamics in the mode space}
In order to clarify the difference between the dynamics of the present
model and that of FPU-type systems, we have investigated the evolution 
of the low-frequency Fourier modes. In Fig.~\ref{modes} we have reported
the power spectra of some long-wavelength modes. For the sake of comparison, 
the same quantities are reported for a diatomic FPU-chain, that is
characterized by an anomalous transport. At variance with the 
FPU model, in the rotators there is no sharp peak (which is a signal of an 
effective propagation of correlations \cite{Lepri2}). 
Quite differently, the low-frequency part of the spectrum is described very
well by a Lorentzian with halfwidth $\gamma = D k^2$ ($D \approx 4.3$). 
This represents an independent proof that energy diffuses, as one expects 
whenever the Fourier law is established.

\begin{figure}[h]
\psfig{file=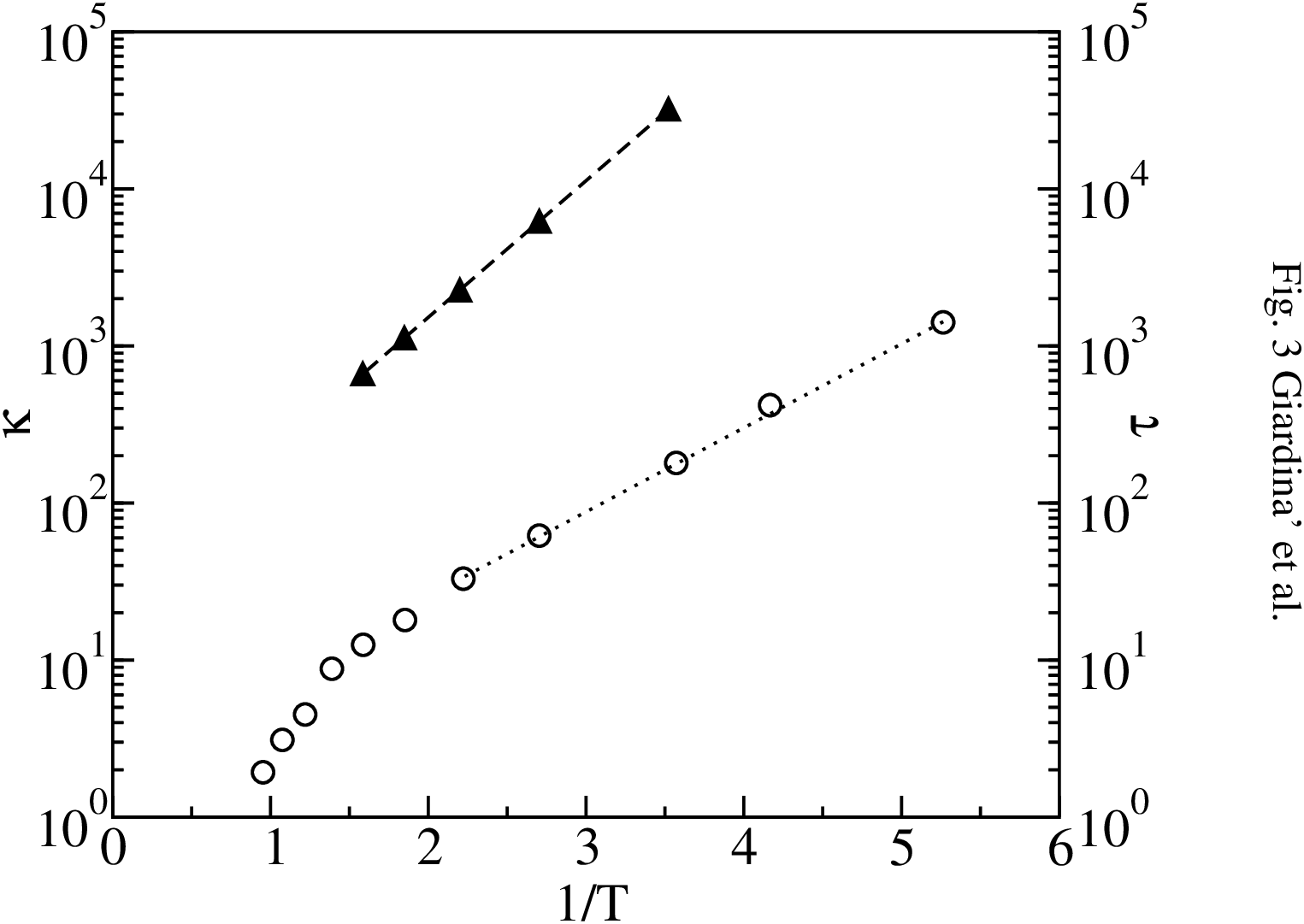,width=8.5 truecm,angle=0}
\caption{Thermal conductivity $\kappa$ (labels on the left axis) versus the 
inverse temperature $1/T$ in the rotator model (open circles). Triangles
correspond to the average time separation between consecutive phase jumps 
(labels on the right axis) in the same system.} 
\label{cond2}
\end{figure}

\paragraph
{Temperature dependence of the thermal conductivity}
The most natural question arising from these results concerns the reason
for the striking difference with other symmetric models such as the 
FPU-$\beta$ system. As long as each $(q_{i+1}-q_{i})$ remains confined to 
the same valley of the potential, there cannot be any qualitative
difference with the models previously studied in the literature. Jumps
thorugh the barrier, however, appear to act as localized random kicks that 
contribute to scattering the low-frequency modes and thus to a finite
conductivity. If this intuition is correct, one should find analogies 
between the temperature dependence of the conductivity and the 
average escape time from the potential well. To this aim, we have computed 
$\kappa$ for different temperature values by performing microcanonical
simulations with various energy densities. From the data reported in 
Fig.~\ref{cond2}, one can notice a divergence for $T \to 0$ of the type 
$\kappa \approx \exp (\alpha/T)$ 
with $\alpha \approx 1.2$. An even more convincing evidence of this
behavior is provided by the temperature dependence of the average escape 
time (see the triangles in Fig.~\ref{cond2}) with an exponent 
$\alpha \approx 2$. The latter behaviour can be explained by assuming that 
the jumps are the results of activation processes. Accordingly, the probability
of their occurrence is proportional to $\exp(-\Delta V/T)$, where
$\Delta V$ is the barrier height and the Boltzmann constant is fixed equal
to 1 (as implicitely done throughout the Letter). Since $\Delta V = 2$, 
the whole interpretation is consistent. Moreover, in the absence 
of jumps, the dependence of the conductivity on the length should be the
same as in FPU-systems, i.e. $\kappa \approx N^{2/5}$. Therefore, a 
low-frequency mode travelling along the chain should experience a 
conductivity of the order of ${\overline N}^{2/5}$, where ${\overline N}$ 
is the average separation between jumps. Under the assumption of
a uniform distribution of phase jumps, their spatial separation is of 
the same order of their time separation, so that we can
expect that $\kappa \approx \exp[2\Delta V/(5T)]$. On the one hand, this
heuristic argument explains why and how such jumps contribute to a normal 
transport. On the other hand, the numerical disagreement between the observed
and the expected value of the exponent $\alpha$ (1.2 vs. 0.8) indicates 
that our analysis needs refinements. In fact, we should, e.g., notice that, 
in the low-energy limit, nonlinearities become negligible, implying that 
deviations from the asymptotic law $\kappa \approx N^{2/5}$ should become
relevant.

\paragraph
{Further checks}
In order to test the conjecture that jumps are responsible for a normal
heat transport, we have decided to investigate some other models.
First, we have considered a double-well potential 
$V(x) = -x^2/2 + x^4/4$ (the same as in FPU with a different 
sign for the harmonic term). The results of the direct simulations are 
reported in Fig.\ref{av_flux} (see triangles) for a temperature 
corresponding again to a quarter of the barrier height. The finiteness
of the conductivity ans its numerical value is confirmed by the computation
of $\kappa$ through the Green-Kubo formula (see the light-grey shaded region).

Finally, we have considered an asymmetric version of the rotator model, namely
$V(x) = A - \cos(x) + 0.4 \sin(2x)$, where $A$ is fixed in such a way that 
the minimum of the potential energy is zero, and the temperature again
corresponds to one quarter of the barrier height. In this case too, the 
conductivity is finite, confirming our empirical idea that the jumps are 
responsible for 
breaking the coherence of the energy flux and, in turn, for the 
finite conductivity. However, from the point of view of Ref.~\cite{PC99}, this
result is quite unexpected, since, in view of the model asymmetry, one
expects that the pressure $\phi = \langle \sum f_i \rangle/N$ is non-zero. 
Nonetheless, microcanonical simulations show that although the distribution 
of forces is definitely asymmetric, their average value is numerically 0. 
This can be understood by noticing that in view of the boundedness of
the potential, the system cannot sustain any compression. Therefore, no
contradiction exists with Ref.~\cite{PC99}.

In conclusion, in this Letter we have reported about the first evidence
of normal heat transport in 1d systems with momentum conservation.
Such a behaviour appears to be connected with jumps between neighbouring
potential valleys. From the dynamical point of view, it is natural to
ask what are the peculiar properties of such jumps that
make them so different from other types of nonlinear fluctuations that
may occur in single-well type of potentials. The only clearly distinctive
feature that we have found is the ``hyperbolic'' type of behaviour
in the vicinity of a maximum of the potential which has to be confronted 
with the typical ``ellyptic'' character of the oscillations around the minima.
We hope to be able to understand in the future whether this is truly 
the reason for the difference in FPU and rotator systems.

We thank S. Lepri for several profitable
discussions.  One of us (AP) thanks Gat Omri for some remarks about our
last comments. ISI in Torino is acknowledged for the kind hospitality during
the Workshop on "Complexity and Chaos" 1998, where part of this work was 
performed.


\begin{thebibliography}{10}
\bibitem{note1} Throughout the paper we shall always assume $m_i =1$ unless
otherwise stated.
\bibitem{note2} The lattice spacing is assumed equal to 1
without loss of generality.
\end{thebibliography}
\end{document}